\documentclass[8.5pt,twoside,twocolumn]{article}
\usepackage[super,sort&compress,comma]{natbib} 
\usepackage{mhchem}
\usepackage{times,mathptmx}
\usepackage{sectsty}
\usepackage{balance} 

\usepackage{graphicx} 
\usepackage{lastpage}
\usepackage[format=plain,justification=raggedright,singlelinecheck=false,font=small,labelfont=bf,labelsep=space]{caption} 
\usepackage{fancyhdr}

\usepackage{amssymb}
\usepackage{color}

\begin{document}

\thispagestyle{plain}
\fancypagestyle{plain}{
\fancyhead[L]{\includegraphics[height=8pt]{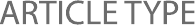}}
\fancyhead[C]{\hspace{-1cm}\includegraphics[height=20pt]{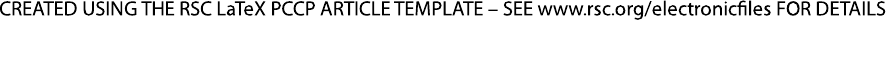}}
\fancyhead[R]{\includegraphics[height=10pt]{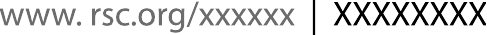}\vspace{-0.2cm}}
\renewcommand{\headrulewidth}{1pt}}
\renewcommand{\thefootnote}{\fnsymbol{footnote}}
\renewcommand\footnoterule{\vspace*{1pt}%
\hrule width 3.4in height 0.4pt \vspace*{5pt}} 
\setcounter{secnumdepth}{5}

\makeatletter 
\def\subsubsection{\@startsection{subsubsection}{3}{10pt}{-1.25ex plus -1ex minus -.1ex}{0ex plus 0ex}{\normalsize\bf}} 
\def\paragraph{\@startsection{paragraph}{4}{10pt}{-1.25ex plus -1ex minus -.1ex}{0ex plus 0ex}{\normalsize\textit}} 
\renewcommand\@biblabel[1]{#1}            
\renewcommand\@makefntext[1]%
{\noindent\makebox[0pt][r]{\@thefnmark\,}#1}
\makeatother 
\renewcommand{\figurename}{\small{Fig.}~}
\sectionfont{\large}
\subsectionfont{\normalsize} 

\fancyfoot{}
\fancyfoot[LO,RE]{\vspace{-7pt}\includegraphics[height=9pt]{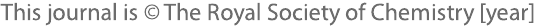}}
\fancyfoot[CO]{\vspace{-7.2pt}\hspace{12.2cm}\includegraphics{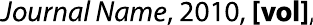}}
\fancyfoot[CE]{\vspace{-7.5pt}\hspace{-13.5cm}\includegraphics{RF}}
\fancyfoot[RO]{\footnotesize{\sffamily{1--\pageref{LastPage} ~\textbar  \hspace{2pt}\thepage}}}
\fancyfoot[LE]{\footnotesize{\sffamily{\thepage~\textbar\hspace{3.45cm} 1--\pageref{LastPage}}}}
\fancyhead{}
\renewcommand{\headrulewidth}{1pt} 
\renewcommand{\footrulewidth}{1pt}
\setlength{\arrayrulewidth}{1pt}
\setlength{\columnsep}{6.5mm}
\setlength\bibsep{1pt}

\twocolumn[
  \begin{@twocolumnfalse}
\noindent\LARGE{\textbf{Manipulating shear-induced non-equilibrium transitions in colloidal films by feedback control}}
\vspace{0.6cm}

\noindent\large{\textbf{Tarlan~A.~Vezirov,\textit{$^{1}$} Sascha~Gerloff,\textit{$^{2}$} and
Sabine~H.~L.~Klapp\textit{$^{2}$}}}\vspace{0.5cm}

\noindent\textit{\small{\textbf{Received Xth XXXXXXXXXX 20XX, Accepted Xth XXXXXXXXX 20XX\newline
First published on the web Xth XXXXXXXXXX 200X}}}

\noindent \textbf{\small{DOI: 10.1039/b000000x}}
\vspace{0.6cm}

\noindent \normalsize{Using Brownian Dynamics (BD) simulations we investigate non-equilibrium transitions of sheared colloidal films under controlled shear stress $\sigma_{\mathrm{xz}}$. In our approach the shear rate 
$\dot\gamma$ is a dynamical variable, which relaxes on a timescale $\tau_c$ such that 
the instantaneous, configuration-dependent stress $\sigma_{\mathrm{xz}}(t)$ approaches a pre-imposed value.
Investigating the dynamics under this ''feedback-control'' scheme we find unique behavior in regions where the flow curve $\sigma_{\mathrm{xz}}(\dot\gamma)$ of the uncontrolled system is monotonic. However, in non-monotonic regions our method allows to {\em select} between dynamical states characterized by different in-plane structure and viscosities. Indeed, the final state strongly depends on $\tau_c$ relative to an {\em intrinsic} relaxation time of the uncontrolled system. The critical values of $\tau_c$ are estimated on the basis of a simple model.}
\vspace{0.5cm}
 \end{@twocolumnfalse}
  ]

\footnotetext{\textit{$^{1}$~Institut f\"ur Theoretische Physik,
  Hardenbergstr. 36,
  Technische Universit\"at Berlin,
  D-10623 Berlin,
  Germany,
  E-mail: tarlan.a.vezirov@tu-berlin.de}}
\footnotetext{\textit{$^{2}$~Institut f\"ur Theoretische Physik,
  Hardenbergstr. 36,
  Technische Universit\"at Berlin,
  D-10623 Berlin,
  Germany}}

\section{Introduction\label{SEC:INTRO}}
Soft matter under shear flow can display rich nonlinear behaviour involving transitions between different dynamical states \cite{Ripoll08,Strehober13}, spontaneous spatial symmetry-breaking \cite{Das05}, shear-banding \cite{Moorcroft13,Poon,Fielding11,Lu00}
(for recent reviews, see Refs.~\cite{Callaghan08, Briels08, Manneville08,Olmsted08,Fielding_Review}), rheochaos \cite{Aradian05,Aradian06,Lootens03} and heterogeneous local dynamics \cite{Zausch09,Chaudhuri13}. These intriguing phenomena often strongly affect the {\it rheological} properties of the system. 
Understanding shear-induced effects in, e.g., complex surfactant solutions \cite{Cates06} or liquid crystals \cite{Ripoll08,Strehober13}, colloids \cite{Besseling12,Derks09,Vezirov13}, soft glasses \cite{Zausch09,Chaudhuri13,Coussot}, and active fluids \cite{Cates08}, is thus a major current topic connecting non-equilibrium physics and soft material science. 

A quantity of particular interest is the flow (or constitutive) curve \cite{Lu00,Fielding_Review}, that is, the shear stress $\sigma$ as function of the shear rate $\dot\gamma$, both of which can serve as experimental control parameters. 
In many examples, the curve $\sigma(\dot\gamma)$ behaves not only nonlinear (indicating shear-thinning \cite{Burrell,Cheng}, shear-thickening \cite{Cheng,Fall},
sometimes connected irregular (chaotic) rheological response \cite{Lootens03,Bandy01}), but becomes also multivalued, i.e. different shear rates lead to the same stress. 
In complex fluids of e.g. wormlike micelles, this multivalued property is in fact a universal indicator of a shear-banding instability, 
specifically, gradient banding, where the (formerly homogeneous) system separates in gradient direction into coexisting bands characterized by a smaller and a larger local shear rate \cite{Fielding_Review} (note that this is different from the more exotic vorticity banding, i.e., the formation of bands along the vorticity direction as discussed e.g., in
Refs.~\cite{Olmsted08,Fielding_Review,Kang}). 
In soft (colloidal) glasses, multivalued functions $\sigma(\dot\gamma)$ occur as transient phenomena after a sudden switch-on of shear stress (Bauschinger effect) \cite{Frahsa}, or in the vicinity of the so-called yield stress \cite{Varnik04}; 
in these systems one observes strong dynamical heterogeneities
\cite{Chaudhuri13}. A further intriguing feature is that close to such nonmonotonicities, the system's behaviour strongly depends on whether one uses the shear stress or the shear rate as a control parameter \cite{Aradian06,Hess10}. In fact, both choices are common in rheological experiments \cite{Coussot,Hu98,Radraraju}.

Here we present BD computer simulation results of yet another system with multivalued stress-shear curve, that is, a thin colloidal film sheared from an (equilibrium) state within in-plane crystalline order in a planar Couette geometry. 
As shown in previous experimental \cite{Palberg} and simulation \cite{Vezirov13,Messina06} studies, such films can display successive non-equilibrium transitions from square crystalline over molten into hexagonal states. Here we demonstrate that the structural transitions lead again to non-monotonic flow curves, with a shape reminding that of materials which perform a solid-to-liquid transition beyond a critical (yield) stress \cite{Fang05}.

Based on this nonlinear behaviour, we then investigate the films in presence of controlled shear stress. In fact, so far most simulations of sheared colloids have been done under fixed shear rate 
$\dot\gamma$, exceptions
being e.g. Refs.~\cite{Chaudhuri13,Varnik04,Mansard13},  where constant $\sigma$ has been realized by fixing the force acting on the atoms forming the walls.
Here we introduce an alternative, easy-to-apply scheme to control $\sigma$ which has been previously used by us in continuum approaches of sheared liquid crystals \cite{Hess10}. In that scheme 
$\dot\gamma$ (which directly enters the BD equations of motion) becomes a dynamical variable whose time dependence is governed by a relaxation equation involving a time scale
$\tau_c$. The relaxation
is based on the difference between the instantaneous (configuration-dependent) stress $\sigma(t)$ and a preimposed value $\sigma_0$. The idea of adapting the shear rate $\dot\gamma$ is inspired by experiments of stress-controlled systems \cite{Hu98, Bonn11}. Our scheme differs from earlier schemes \cite{Chaudhuri13,Varnik04,Mansard13} where the desired value $\sigma_0$ is imposed instantaneously.
Moreover, due to the coupling to the particle positions our method corresponds to a ''feedback'' (closed-loop) control scheme, which is similar in spirit to e.g. a Berendsen thermostat for temperature control \cite{Berendsen}. However, here the choice for $\tau_c$ is found to be crucial for the observed dynamical behaviour. In particular we demonstrate that,
if our scheme is applied within the multivalued regime of $\sigma(\dot\gamma)$, the final state strongly depends 
on the magnitude of  $\tau_c$ relative to important intrinsic time scales
of the system. Thus, the stress-control can be used to deliberately {\em select} a desired dynamical state.

\section{Model and simulation details}
\label{SEC:MODS}
We consider a model colloidal suspension where two particles with distance $r_{ij}$ interact via a combination of a repulsive Yukawa potential, 
$u_{\mathrm{Yuk}}(r_{ij}) = W\exp\left[-\kappa r_{ij}\right]/r_{ij}$ and a repulsive soft-sphere potential $u_{\mathrm{SS}}(r_{ij})=4\epsilon_{\mathrm{SS}}(d/r_{ij})^{12}$ involving the particle diameter $d$ \cite{Vezirov13}. The interaction parameters are set in accordance to a real 
suspension of charged silica macroions [for details, see ref. \cite{Klapp08b,Klapp08a}]. Specifically, at the density considered (see below),
the interaction strength $W/(k_{\mathrm{B}}Td)\approx 123$ (where $k_{\mathrm{B}}T$ the thermal energy) and the inverse screening length $\kappa\approx 3.2d$.
Spatial confinement is modeled by two plane parallel, smooth, uncharged surfaces 
separated by a distance $L_{\mathrm{z}}$ along the $z$ direction and of infinite extent in the $x$--$y$ plane. We employ a purely repulsive fluid-wall decaying as $z^{-9}$ [see ref. \cite{Vezirov13}]. This is motivated by previous investigations of the equilibrium layer structure, where we found a very good agreement with experiments \cite{Klapp08a, Klapp08b}. 

Our investigations are based on standard BD simulations in three dimensions, where the position of particle $i$
is advanced according to \cite{Ermak75}
\begin{equation}
\mathbf{r}_{i}(t+\delta t) = \mathbf{r}_{i}(t)+\frac{D_{0}}{k_{B} T}\mathbf{F}_{i}(t)\delta t+\delta \mathbf{W}_{i}+\dot{\gamma}z_{i}(t)\delta t \mathbf{e}_{x},
\label{EQ:Eqmot}
\end{equation}
where $\mathbf{F}_{i}$ is the total conservative force acting on particle $i$.
Within the framework of BD, the influence of the solvent on each colloidal particle is mimicked by a friction constant and a random Gaussian displacement $\delta \mathbf{W}_{i}$. The friction constant is set to $(D_{0}/k_{B} T)^{-1}$, where $D_{0}$ is the short-time diffusion coefficient. The value $\delta \mathbf{W}_{i}$ has zero mean and variance $2D_{0}\delta t$ for each Cartesian component. The timescale of the system was set to $\tau$=$\sigma^{2}/D_{0}$, which defines the so-called Brownian time.
We impose a linear shear profile [see last term in eqn.~(\ref{EQ:Eqmot})] representing flow in $x$- and gradient in $z$-direction, characterized by uniform shear rate $\dot{\gamma}$. We note that, despite the application of a linear shear profile, the real, steady-state flow profile can be nonlinear \cite{Delhom}.
This approximation has also been employed in other recent simulation studies of sheared colloids \cite{Besseling12,Lander13, Rottler07}; the same holds
for the fact that we neglect hydrodynamic interactions. 

One quantity of prime interest in our study is the $x$-$z$ component of the stress tensor, 
\begin{equation}
\label{stress}
\sigma_{\mathrm{xz}} =\left\langle \frac{1}{V}\sum_{i}\sum_{j>i}F_{x,ij}z_{ij}\right\rangle.
\end{equation}
Thus, we consider only the configuration-dependent contribution to $\sigma_{\mathrm{xz}}$; the kinematic contribution (which involves the velocity components in $x$- and $z$-direction) is negligible under the highly confined conditions here. 
We note that, apart from the kinematic contribution, eqn.~(\ref{stress}) also neglects higher-order contributions involving gradients\cite{Irving,Evans}. In continuum approaches, one typically includes non-local terms which are essential for the description of interfaces between shear-bands \cite{Aradian06, Lu00}. The importance of such terms in our highly confined system, which is characterized by pronounced layer formation, remains to be investigated.
\\

Based on the shear stress, we introduce a feedback scheme as follows. Starting from an initial value for $\dot\gamma$ we calculate, in each time step, 
the configuration-dependent stress $\sigma_{\mathrm{xz}}$ from eqn.~(\ref{stress})
and adjust $\dot\gamma$ via the relaxation equation 
\begin{equation}
\label{Control}
\frac{d}{dt}\dot\gamma=\frac{1}{\tau_c}\frac{\sigma_0-\sigma_{\mathrm{xz}}(t)}{\eta_0},
\end{equation}
where $\sigma_0$ is a {\em pre-imposed} value of $\sigma_{\mathrm{xz}}$, and $\tau_c$ determines the time scale of relaxation. Also, $\eta_0$ is the shear viscosity obtained for $\dot\gamma\to 0$ 
(Newtonian regime). This control scheme is inspired by experiments under fixed stress [see, e.g. ref. \cite{Hu98}], where the adaptation of the shear rate to a new stress value always takes a finite time. 

From a more formal point of view, we note that through eqn.~(\ref{Control}), $\dot\gamma$ becomes an additional dynamical variable. Therefore, and since $\sigma_{\mathrm{xy}}(t)$ depends on the instantaneous configuration $\{{\bf r}_i(t)\}$ of the particles, simultaneous solution
of the $N+1$ equations of motion (\ref{EQ:Eqmot}) and (\ref{Control}) forms a closed feedback loop with {\em global} coupling. 
Interestingly, this feedback scheme is in accordance with the common view that, in a stable system, the shear stress $\sigma_{\mathrm{xz}}$
should {\it increase} with the applied shear rate. This can be shown (at least for a homogeneous system) by a linear stability analysis of eqn.~(\ref{Control})
as outlined in Appendix~A.

In our numerical calculations, we focus on systems at high density, specifically $\rho d^3=0.85$,
and strong confinement, $L_{\mathrm{z}}=2.2d$. The corresponding equilibrium system
forms a colloidal bilayer with crystalline in-plane structure \cite{Klapp08a}. We also show some data with $L_{\mathrm{z}}=3.2d$, yielding three layers. The values
$L_{\mathrm{z}}=2.2$ and $3.2$ have been chosen
because the equilibrium lattice structure is square \cite{Klapp08a} (other values would lead to hexagonal equilibrium structures which 
do not show the shear-induced transitions discussed here). The number of particles
was set to $N=1058$ and $1587$, the width of the shear cell to $L\approx 23.8d$ and $24.2d$ for  $L_{\mathrm{z}}=2.2d$ and $3.2d$, respectively. Periodic boundary conditions were applied in flow ($x$) and vorticity ($y$) direction. The timestep was set to $\delta t=10^{-5}\tau$ where $\tau=d^{2}/D_{0}$ the time unit. 

The system was equilibrated for $10\times10^{6}$ steps (i.e. 100$\tau$~). Then the shear force was switched on. After the shearing was
started the simulation was carried out for an additional period of 100$\tau$. During this time the steady state was reached. Only after these preparations we started to analyze the considered systems.

\section{Shear-induced transitions\label{SEC:STRESS}}
\begin{figure}
\includegraphics[width=0.8\linewidth]{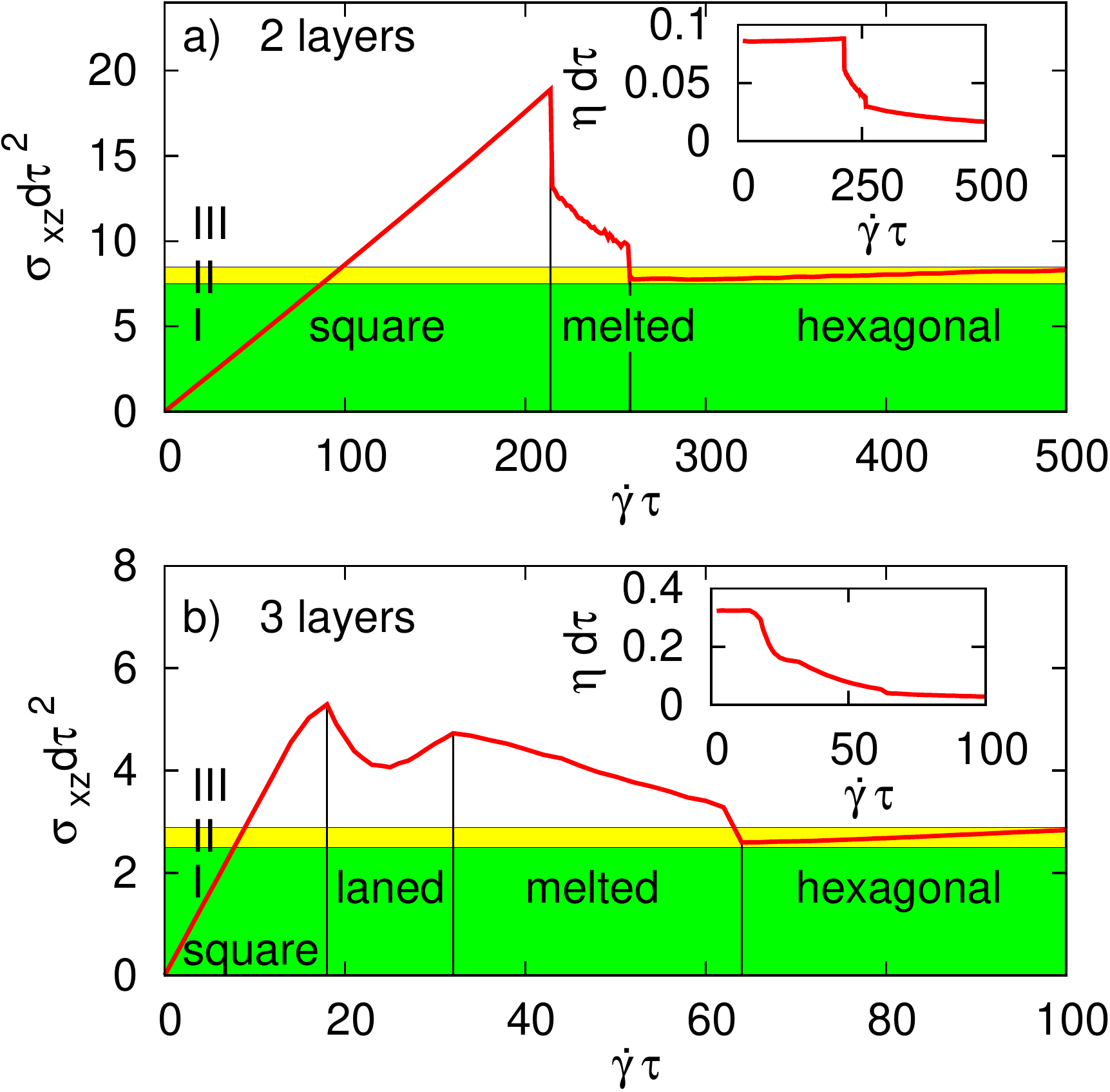}
\caption{(Color online) Steady state shear stress and shear viscosity (insets) for bi- and trilayer systems as function of the applied shear rate. Regions indicated as I, II, III are discussed in the main text. }
\label{FIG:Viscosity}
\end{figure}

\begin{figure}
\includegraphics[width=\linewidth]{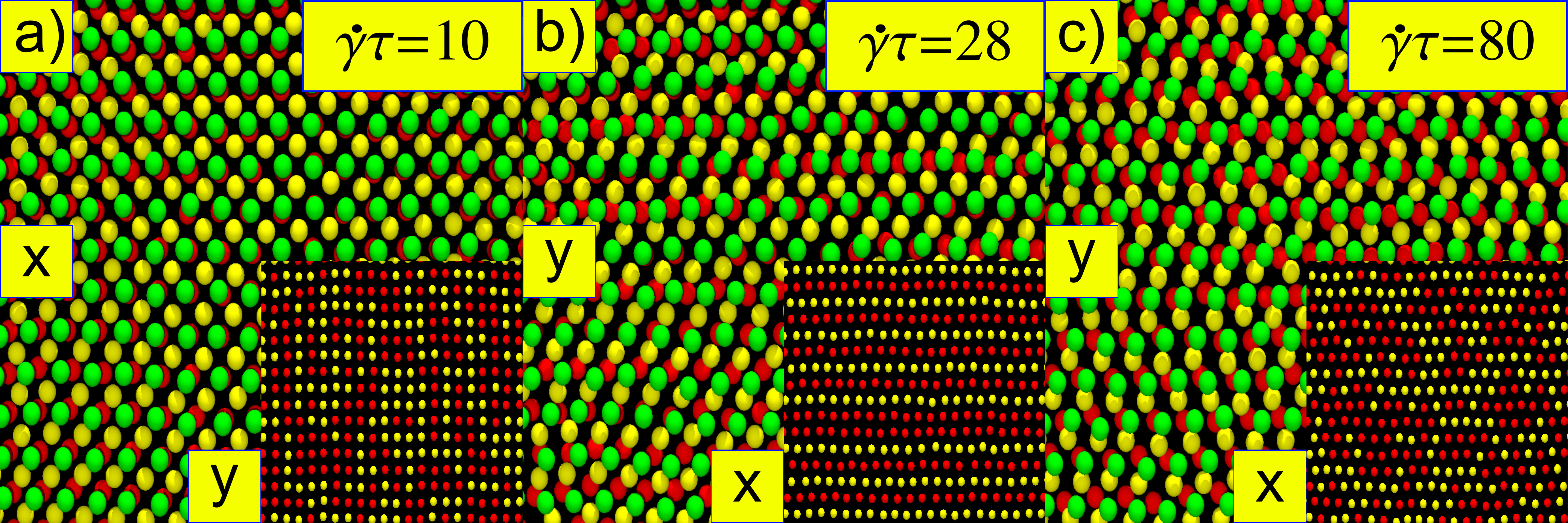}
\caption{(Color online) Snapshots of a colloidal trilayer at three different shear rates corresponding to (a) square, (b) laned, and (c) hexagonal state. 
In the main figures, the three colours correspond to particles in the three layers. In the insets, the two colours indicate the particles of the middle layer which are positioned at $z>0$ (red) and $z<0$ (yellow), respectively (where $z=0$ is the middle plane).}
\label{FIG:Snapshot}
\end{figure}

We start by considering flow curves for systems at constant $\dot\gamma$. The functions $\sigma_{\mathrm{xz}}(\dot\gamma)$ for both, two- and three layer systems
are plotted in Fig.~\ref{FIG:Viscosity}, where we have included data for the viscosities $\eta = \sigma_{\mathrm{xz}}/\dot{\gamma}$. Note that we have defined
$\eta$ via the externally applied rate $\dot\gamma$ rather than via the {\it effective} shear rate within the system (characterized by the average velocity
of the layers \cite{Vezirov13}), which can show significant deviations from $\dot\gamma$.
As seen from Fig.~\ref{FIG:Viscosity}, both the bi- and the trilayer systems are characterized by a nonmonotonic flow curve, accompanied by pronounced shear-thinning. At small shear rates, the systems
display linear stress, related to an Newtonian response of the square in-plane lattice structure. 
An exemplary simulation snapshot for the three-layer system is shown in Fig.~\ref{FIG:Snapshot}a
[see ref.~\cite{Vezirov13} for corresponding results for the bilayer]. 
In fact, within the Newtonian regime (square state) the layer velocities are zero, i.e., the particles are ''locked'' \cite{Vezirov13}. In this state the lattice structure persists and the system reacts to the displacement of the particles in elastical manner.
Further increase of the shear rate then destroys the square order. In the bilayer, the system then enters a "shear-molten" state characterized by the absence of translational order within the layers
(as indicated, e.g., by a non-zero long-time diffusion constant in lateral directions \cite{Vezirov13}).
At the same time, the entire system starts to flow, that is,
the layer velocities increase from zero to non-zero values \cite{Vezirov13}. From the function $\sigma_{\mathrm{xz}}(\dot\gamma)$ plotted in Fig.~\ref{FIG:Viscosity}a, the appearance of the shear-molten state is indicated
by a sudden decrease of $\sigma_{\mathrm{xz}}$, implying the onset of shear-thinning. In fact, with the shear-molten regime the slope of the flow curve is negative. For bulk systems, such a behavior
implies that the system is mechanically unstable \cite{Tanner, Olmsted}. Here we are considering a strongly confined system, where the macroscopic arguments cannot be immediately applied.
Nevertheless, it is an interesting question to which extent the flow curve pertains to a true steady state in the regime where the shear rate has values corresponding to shear-molten configurations.
Investigating the shear stress as function of strain (see Appendix~B) we find that 
$\sigma_{\mathrm{xz}}$ assumes indeed a constant value on the time scale of our simulations; however, the relaxation time is extremely long (see next section).
We also mention that our observation of shear-molten (long-time) configurations
in the regime, where $\sigma_{\mathrm{xz}}$ decreases with $\dot\gamma$, is consistent 
with findings in an earlier theoretical study of a colloidal suspension under shear \cite{Harrowell}. 

Somewhat different behaviour is found in the trilayer system which displays, before melting, 
an intermediate state
[see Fig.~\ref{FIG:Snapshot}b]: This state is characterized by a non-zero layer velocity. In addition, the middle layer separates into two sublayers, in which the particles are ordered in "lanes" [see inset of Fig.~\ref{FIG:Snapshot}b] 
and move with the velocity of the corresponding outer layer (a more detailed discussion of this ''laned'' state will be given elsewhere \cite{Vezirov14}). 
Only further increase of $\dot\gamma$ then yields
a shear-molten state characterized by a decreasing flow curve (in analogy to the bilayer).
Finally, both systems transform into a state with in-plane hexagonal ordering [see Fig.~\ref{FIG:Snapshot}c] and low viscosity. 
In this hexagonal state the layer velocity is non-zero \cite{Vezirov13}, that is, the systems flows.
As demonstrated earlier by us
\cite{Vezirov13} the mechanism of relative motion involves collective oscillations of the particles around lattice sites, consistent with recent experiments of 3D sheared colloidal crystals \cite{Derks09}. Regarding the stress, we see that the hexagonal regime is (in both systems) characterized by a slight increase of $\sigma_{\mathrm{xz}}$ with $\dot\gamma$. As a consequence, there is a parameter range (indicated as region "II" in Fig.~\ref{FIG:Viscosity}) where the flow curve is {\em multivalued}, that is, different $\dot\gamma$ lead to the same $\sigma_{\mathrm{xz}}$. In many contexts (such as for worm-like micells), multivalued flow curves are associated with the phenomenon of shear-banding, that is, the separation of the system into spatial regions with different shear rates. In our case, where the system consists of two or three layers such a separation can not occur. Instead, the non-monotonic stress curve is a consequence of the structural transitions of the system induced by the shear.

\section{Intrinsic time scales}
\label{SEC:TS}
\begin{figure}
\includegraphics[width=0.9\linewidth]{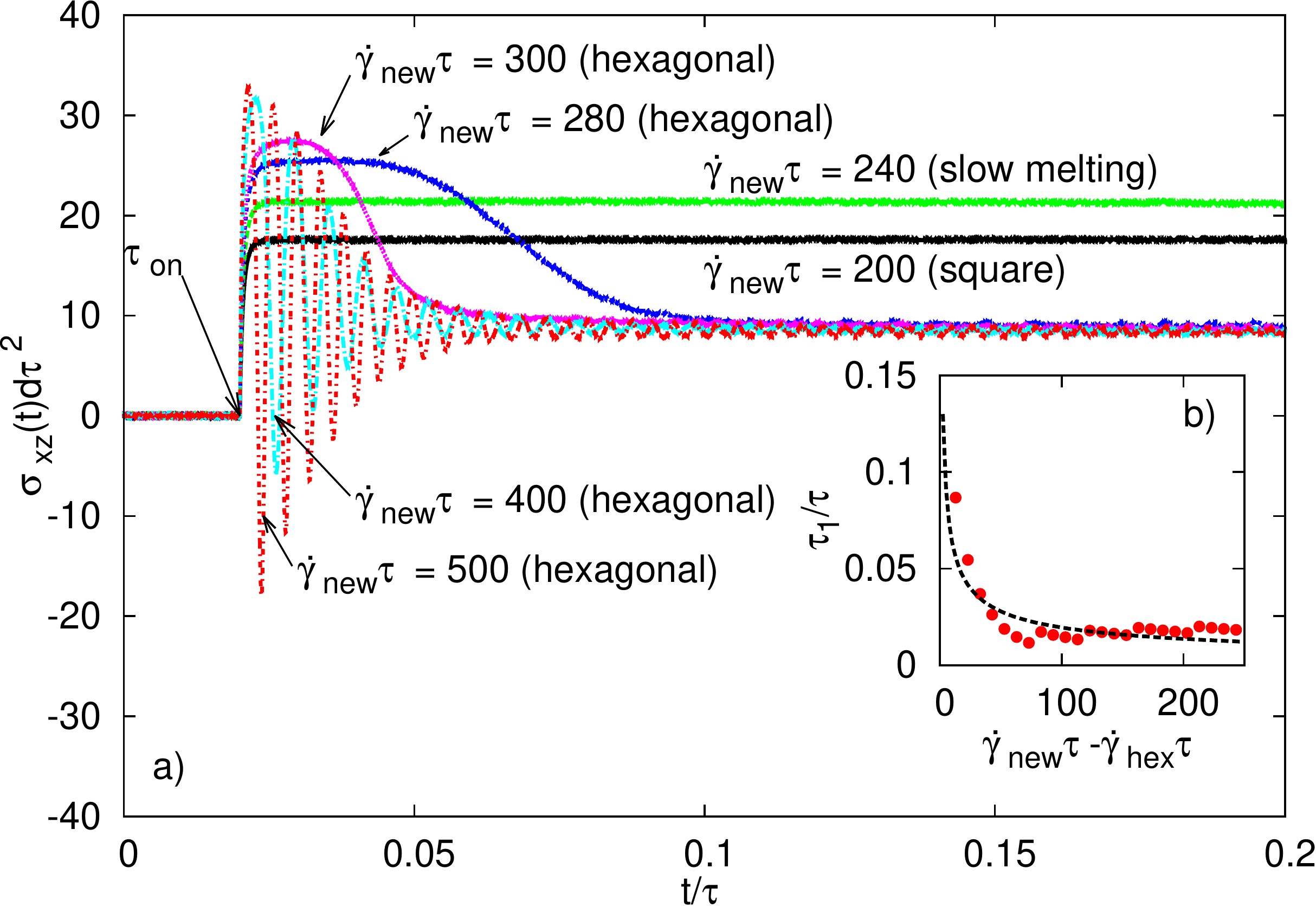}
\caption{(Color online) Response of $\sigma_{\mathrm{xz}}(t)$ to a sudden switch-on (at time $\tau_{\mathrm{on}}$) of shear with different rates $\dot\gamma_{\mathrm{new}}\tau$. The simulations were started from the equilibrium (square) state in a bilayer system. The inset shows the fit of the relaxation times $\tau_1$ according to eqn.~(\ref{intrinsic}).}
\label{FIG:Relaxation1}
\end{figure}

Before exploring the impact of shear-stress control, which involves a time scale itself through the parameter $\tau_c$ [see eqn.~(\ref{Control})], 
we take a closer look at the {\em intrinsic} time scales characterizing the uncontrolled systems.
We focus on the bilayer (the same findings apply qualitatively on the trilayer) and consider
the response of the unsheared equilibrium system, which is in a square state, to a {\em sudden} switch-on 
at time $\tau_{\mathrm{on}}$ of shear with rate $\dot\gamma_{\mathrm{new}}$. 
The resulting time dependence of the instantaneous stress is plotted in Fig.~\ref{FIG:Relaxation1}.  

If $\dot\gamma_{\mathrm{new}}\tau$ has a value pertaining to the square state, the shear stress jumps at $\tau_{\mathrm{on}}$ to a non-zero value but then settles quickly to its steady-state value [see Fig.~\ref{FIG:Viscosity}]. At shear rates corresponding to the shear-molten state we can also observe a relaxation at some non-zero value. However, it should be emphasized 
that this value is {\it transient} in character. The true, steady state value is only achieved at much longer simulation times
(see the stress-strain relations presented in Appendix~B).
\begin{figure}[h]
\includegraphics[width=0.9\linewidth]{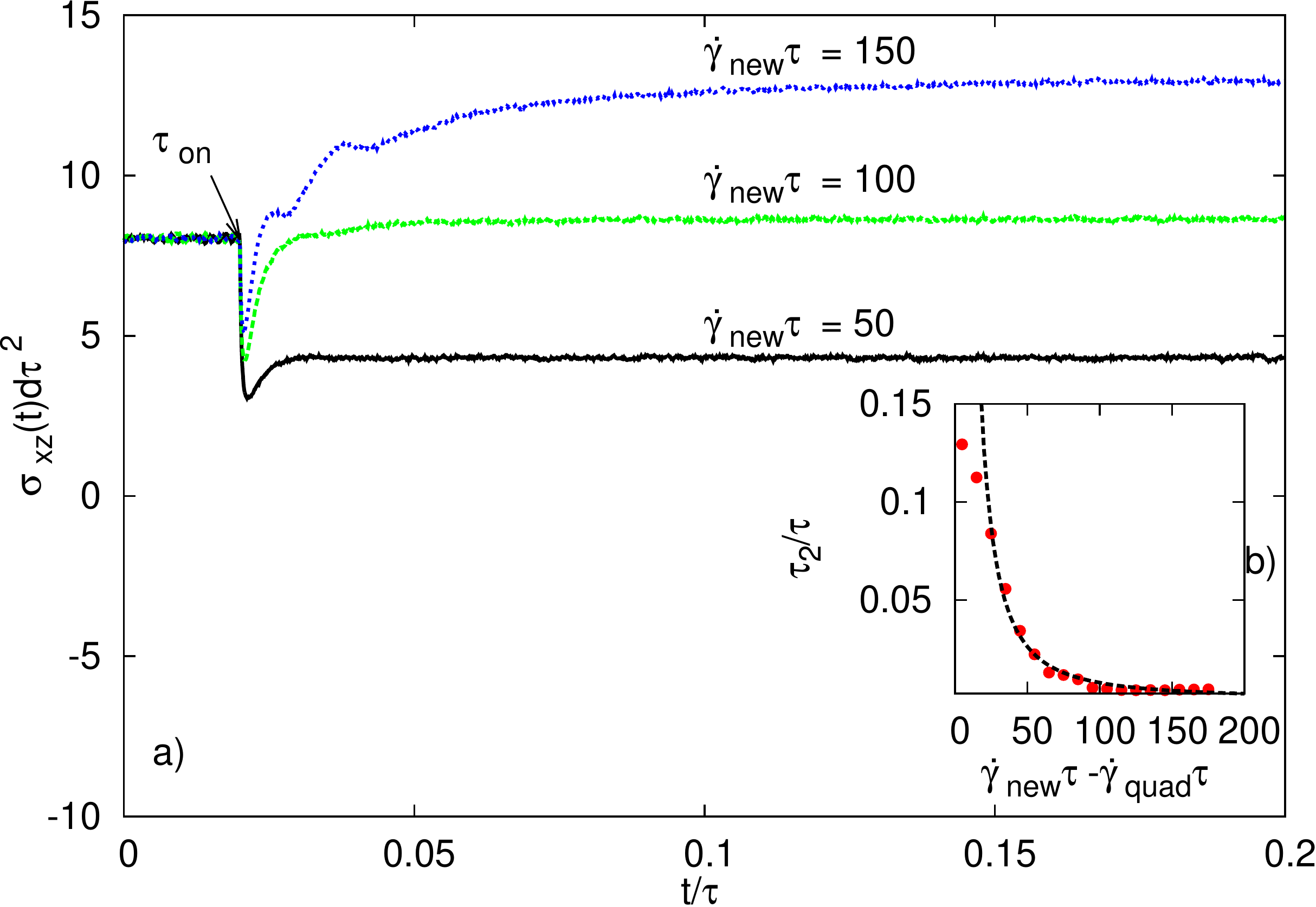}
\caption{(Color online) Response of $\sigma_{\mathrm{xz}}(t)$ to a sudden change (at time $\tau_{\mathrm{on}}$) of shear. The new shear rates  $\dot\gamma_{\mathrm{new}}\tau$  result in a relaxation in the square state. The simulations were started from the hexagonal steady state at $\dot\gamma\tau=400$ in a bilayer system. The inset shows the fit of the relaxation times $\tau_3$ according to eqn.~(\ref{intrinsic}).}
\label{FIG:Relaxation2}
\end{figure}
Finally, for shear rates related to the hexagonal steady state
($\dot\gamma_{\mathrm{new}}\tau > \dot\gamma_{\mathrm{hex}}\tau\approx 257$, [see Fig.~\ref{FIG:Viscosity}]), we 
observe a well-pronounced stress overshoot, similar to what is observed e.g. in soft glassy systems \cite{Frahsa}, wormlike micelles \cite{Padding05} and polymer melts \cite{Delay06}.
Closer inspection shows that the actual value of $\tau_1$ as well as the functional behavior of $\sigma_{\mathrm{xz}}(t)$ strongly depends on the distance between $\dot\gamma_{\mathrm{new}}\tau$ and the threshold
between shear-molten and hexagonal state, $\dot\gamma_{\mathrm{hex}}\tau$: the smaller this distance is, the larger becomes $\tau_1$, and the more sensitive it is against small changes of the shear rate. Moreover, a sudden quench {\em deep} into the hexagonal state leads to an {\em oscillatory}
relaxation of the stress $\sigma_{\mathrm{xz}}(t)$ [see curves $\dot\gamma_{\mathrm{new}}\tau=400$, $500$], with $\tau_1$ (which now corresponds to the relaxation time of the
envelope) being still quite large. Taken together, for $\dot\gamma_{\mathrm{new}}\tau > \dot\gamma_{\mathrm{hex}}\tau$, $\tau_1$ can be fitted according to (see inset
in Fig.~\ref{FIG:Relaxation1})
\begin{equation}
\tau_i=\frac{a_i}{|\dot{\gamma}_{\mathrm{new}}\tau-\dot{\gamma}_i\tau|^{b_i}},
\label{intrinsic}
\end{equation}
where we find $a_1/\tau=0.21$, $b_1=0.52$ (setting $\dot\gamma_1\tau=\dot\gamma_{\mathrm{hex}}\tau$). The oscillations occurring at large $\dot\gamma_{\mathrm{new}}$
induce yet a different time scale 
$\tau_{os}$, which is smaller than $\tau_1$. Specifically, we find $\tau_{\mathrm{os}}/\tau={\cal O}(10^{-3})$. The physical reason for these oscillations is the ''zig-zag" motion of particles in adjacent layers
\cite{Vezirov13}. This motion is accompanied by periodic variations of nearest-neighbor distances, and thus, pair forces, which eventually leads to oscillations of $\sigma_{\mathrm{xz}}(t)$.

Furthermore it is interesting to relate the relaxation times emerging from Fig.~\ref{FIG:Relaxation1} to the structural transition from square to hexagonal state in Fig.~\ref{FIG:Viscosity}. To this end we consider the Peclet number $Pe\equiv\dot\gamma\tau_{\mathrm{Pe}}$ where $\tau_{\mathrm{Pe}}$ is a ``typical'' relaxation time \cite{Stevens}. Identifying $\tau_{\mathrm{Pe}}$ with $\tau_{1}$ and considering shear rates $\dot\gamma$ close to the transition from the quadratic into the shear-molten state, we find  $Pe={\cal O}(10^{0})$. In other words, the shear-induced structural transitions happen at  $Pe\geqq1$, consistent with our expectations.\cite{Stevens} 

For comparison we have also investigated the reverse situation, where the system is initially in a hexagonal steady state at shear rate $\dot{\gamma}_{\mathrm{init}}=400$. We then suddenly change the shear rate towardts a much smaller value and consider the relaxation towards the square equilibrium state. The corresponding behaviour of $\sigma_{\mathrm{xz}}(t)$ is shown in Fig.~\ref{FIG:Relaxation2}. Again we find that, the smaller the difference $\dot\gamma_{\mathrm{new}}\tau-\dot\gamma_{\mathrm{2}}\tau$ is, the larger $\tau_{2}$ becomes (and the more pronounced is the sensitivity to small changes in $\dot\gamma_{\mathrm{new}}$). The resulting relaxation times can be also fitted via eqn.~(\ref{intrinsic}) with $a_2/\tau=22.58$, $b_2=1.73$ and $\dot\gamma_2\tau=215$, whereby $\dot\gamma_2\tau=\dot\gamma_{sq}\tau$ denotes the threshold between the square and the molten states. The result for this fit is visualized in the inset of the Fig.~\ref{FIG:Relaxation2}.

\section{Impact of feedback control}
\label{SEC:FC}
We now discuss the impact of our shear stress control scheme defined in eqn.~(\ref{Control}). The latter involves the zero-shear viscosity, $\eta_0$, which is estimated from Fig.~\ref{FIG:Viscosity} 
to $\eta_0=0.086/d\tau$ and $0.323/d\tau$ for the bilayer and trilayer, respectively.

The overall dynamical behaviour under feedback control strongly depends on the value of
$\sigma_0$ (imposed shear) relative to the flow curve of the original system [see Fig.~\ref{FIG:Viscosity}]. We can differentiate
between regimes~I, II, and III, which are indicated in Fig.~\ref{FIG:Viscosity}.

\begin{figure}
\includegraphics[width=0.95\linewidth]{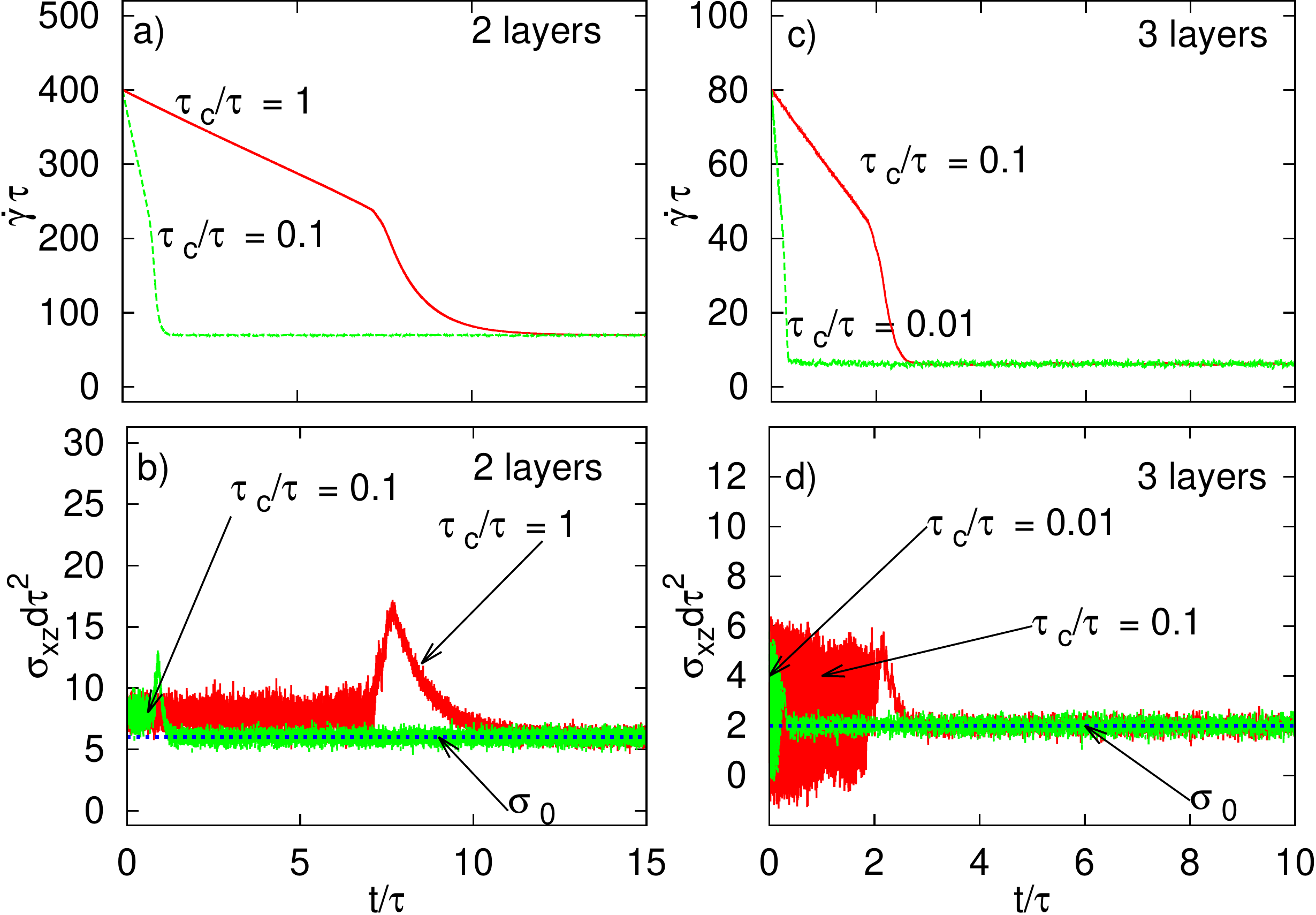}
\caption{(Color online) Time dependence of the instantaneous shear rate and shear stress for a bilayer- [a), b)] and a trilayer system [c), d)] in presence of feedback control within region~I. 
The imposed stress was set to $\sigma_{0}d \tau^2=6(2)$ for the bilayer (trilayer) system. Various values of $\tau_c/\tau$ are considered. The initial configuration is hexagonal.}
\label{FIG:Rel_control1}
\end{figure}

For a $\sigma_0$ chosen in region~I, the response of the system is {\em unique}, that is, the final state is independent of the control timescale $\tau_{c}$, as well as of the initial shear rate $\dot\gamma_{\mathrm{init}}$ and the initial microstructure. Thus, when starting from a square state, with a corresponding initial shear rate $\dot{\gamma}_{\mathrm{init}}$, the system immediately settles at this state. As a more critical test of the injectivity of the flow curve in region I, we plot in Fig.~\ref{FIG:Rel_control1}a and \ref{FIG:Rel_control1}b the functions $\dot{\gamma}(t)$ and $\sigma_{\mathrm{xz}}(t)$ for the bilayer system at $\sigma_0 d\tau^2=6$ and various $\tau_c$, starting from a {\em hexagonal} configuration (and $\dot{\gamma}_{\mathrm{init}}\tau=400$).
In all cases, the shear rate decreases towards the value $\dot{\gamma}\tau\approx 70$ 
and the structure relaxes into the square state pertaining to the value $\sigma_{\mathrm{xz}}d \tau^2=6$ in the {\em uncontrolled} system. This indicates that the square state in region~I is indeed the only fixed point of the dynamics. We also see from Fig.~\ref{FIG:Rel_control1}a that the relaxation time into this steady state increases with $\tau_c$. 
Figure~\ref{FIG:Rel_control1}b additionally shows
that $\sigma_{\mathrm{xz}}(t)$ displays a pronounced peak. The peak indicates the time window in which the initial hexagonal ordering transforms into a square one. In fact, the high values of $\sigma_{\mathrm{xz}}$ at the peak reflect the large friction characterizing the intermediate molten state. Similar behaviour occurs in region~I of the trilayer system [see Fig.~\ref{FIG:Rel_control1}c,d] where, however, fluctuations of 
$\sigma_{\mathrm{xz}}(t)$ are generally larger.

\begin{figure}[h]
\includegraphics[width=0.95\linewidth]{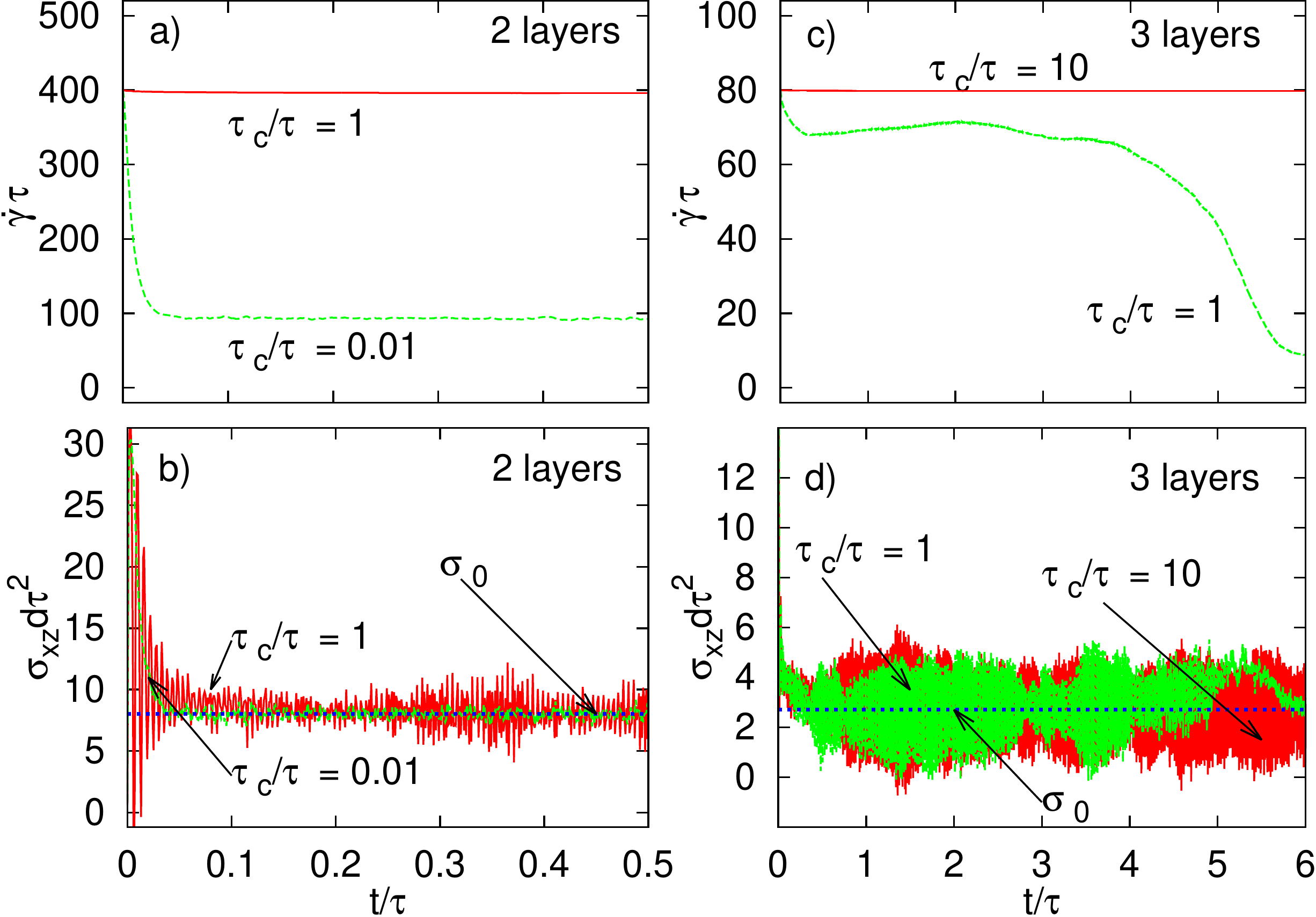}
\caption{(Color online) 
Same as Fig.~\ref{FIG:Rel_control1}, but for  $\sigma_{0}d \tau^2=8(2.7)$ for the bilayer (trilayer) system (region~II). The initial configuration is square.}
\label{FIG:Rel_control}
\end{figure}

\begin{figure}
\includegraphics[width=0.9\linewidth]{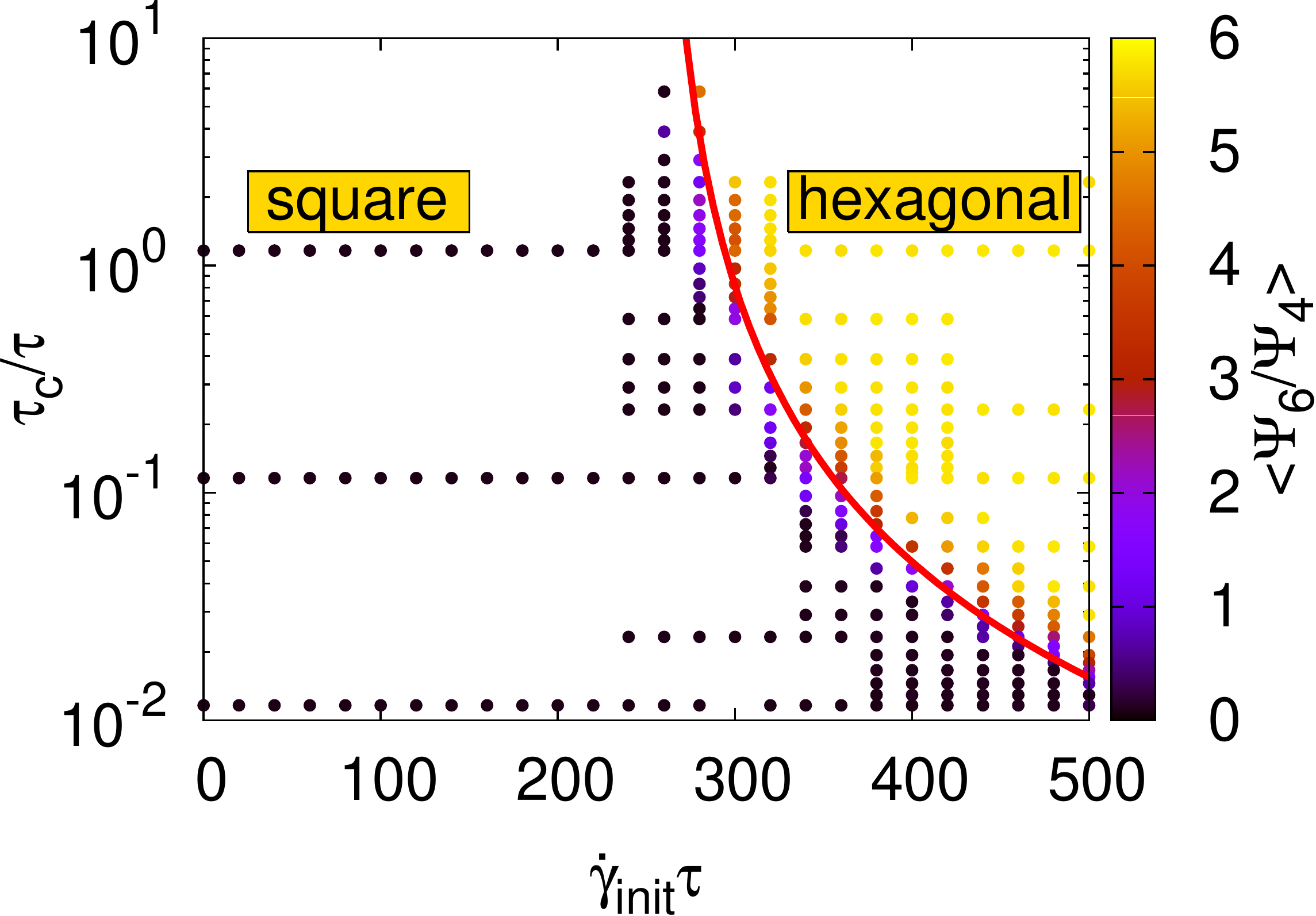}
\caption{(Color online) State diagram indicating long-time lattice structures. 
All simulations were started from a square initial structure and the imposed shear stress was set to $\sigma_0 d\tau^2=8$. 
The line shows the result from eqn.~(\ref{Control_an4}).}
\label{FIG:Heatmap}
\end{figure}

We now choose $\sigma_0$ within region~II of the flow curve, where there are three different shear rates (and thus, three fixed points) pertaining to the same stress [see Fig.~\ref{FIG:Viscosity}]. We focus on systems which are initially in a square configurations, whereas the initial shear rate $\dot{\gamma}_{\mathrm{init}}$ has a value pertaining to the hexagonal state (other initial conditions
will be discussed below).
The impact of $\tau_c$ on the time dependence of $\dot{\gamma}(t)$ and $\sigma_{\mathrm{xz}}(t)$ is shown 
in Fig.~\ref{FIG:Rel_control}. For small values of the control timescale the systems stays in the initial lattice configuration, i.e., 
$\dot\gamma$ relaxes towards the value 
pertaining to the square state ($\dot\gamma\tau\approx 90$). 
Different behaviour occurs at larger values of $\tau_c/\tau$: Although the initial structure is square, the final state is {\em hexagonal}, and the shear rate essentially remains at its high initial value. We stress that these findings crucially depend on the choice of $\dot{\gamma}_{\mathrm{init}}$. In particular, the dependency of the long-time behaviour on $\tau_c/\tau$ only arises for large values of $\dot{\gamma}_{\mathrm{init}}$; for small
values the system remains in the square state irrespective of $\tau_c$. 
An overview of the final dynamical states in the feedback-controlled bilayer at $\sigma_0 d\tau^2=8$ and various combinations
of $\dot{\gamma}_{\mathrm{init}}$ and $\tau_c/\tau$ (assuming a square initial structure) is given in Fig.~\ref{FIG:Heatmap}.
The colour code indicates the ratio of local bond-order parameters $\langle\Psi_6/\Psi_4\rangle$ [for a definition of the $\Psi_n$ see e.g., ref.~\cite{Vezirov13}]. 
The restriction to values $\langle\Psi_6/\Psi_4\rangle\leq 6$ is related to the actual values observed in the simulations. 
From Fig.~\ref{FIG:Heatmap} one clearly sees that for initial shear rates $\dot{\gamma}_{\mathrm{init}}\tau>\dot{\gamma}_{\mathrm{hex}}\tau\approx 257$, the final state of the feedback-controlled system depends
on $\tau_c/\tau$. This is in contrast to the uncontrolled
system which becomes hexagonal for all $\dot{\gamma}_{\mathrm{init}}>\dot{\gamma}_{\mathrm{hex}}$. For a hexagonal initial configuration the diagram (not shown here) looks similar from a qualitative point of view; however, the range of control times where the system 
retains a hexagonal state despite of $\dot{\gamma}_{\mathrm{init}}<\dot{\gamma}_{\mathrm{quad}}$ (with $\dot{\gamma}_{\mathrm{quad}}$ being the threshold between square/molten states) is much smaller. 

We conclude that, by varying $\tau_c$ and the initial structure, we can ''switch''  between the two stable, steady-state configurations arising in the multivalued region of the uncontrolled system. That these states are stable also under feedback (stress) control 
is supported by the linear stability analysis presented in Appendix~A.
Indeed, the dynamics under feedback control {\em never} evolves towards the intermediate, shear-molten states, consistent with the view that these states are mechanically unstable. This holds also in region~III of the flow curve of the uncontrolled system,
e.g., for $\sigma_0 d\tau^2=16(5)$ for the bilayer (trilayer): Here, a small value of $\tau_c$ yields relaxation towards the square state, whereas for large $\tau_c$, the system evolves into a hexagonal state. 
Finally, we note that completely analogous behaviour is found in the trilayer system [see Fig.~\ref{FIG:Rel_control}b,c]
for a $\sigma_0$ pertaining to the regime where square, molten and hexagonal states exist.
\section{Transition line\label{SEC:AST}}
The most significant observation from Fig.~\ref{FIG:Heatmap} is that at high values of $\dot{\gamma}_{\mathrm{init}}$, the feedback-controlled system can achieve
{\em either} the hexagonal or the square configuration, provided that we start from a square configuration and choose $\tau_c/\tau$ accordingly. We now propose a simple model which allows
us to estimate the {\em transition} values of the control time, $\tau_c^{\mathrm{trans}}$. 

The physical idea behind our model is that, with the initial conditions described above, relaxation into the hexagonal state only occurs if the {\em reorganization} time $\tau_{\mathrm{reorg}}$
required by the system to transform from a square into a hexagonal configuration, is smaller than the time $\tau_{\mathrm{decay}}$ in which $\dot\gamma$ decays 
to a value pertaining to the square state. We can estimate $\tau_{\mathrm{decay}}$ from eqn.~(\ref{Control}) if we assume, for simplicity, a {\em linear} relationship
$\sigma_{\mathrm{xz}}(t)=m\dot{\gamma}(t)$ (note that such a relationship is indeed nearly fulfilled {\em within} the square and hexagonal states, see Fig.~\ref{FIG:Viscosity}). Under this assumption eqn.~(\ref{Control}) can be easily solved, yielding 
\begin{equation}
\dot{\gamma}(t)=m^{-1}e^{- m t/ \eta_0 \tau{c}} \left(m \dot{\gamma}_{\mathrm{init}} - \sigma_{0} +  e^{ m t/ \eta_0 \tau{c}} \sigma_{0}\right).
\label{Control_an1}
\end{equation}
From eqn.~(\ref{Control_an1}) we find that the decay time of $\dot\gamma$ to the threshold value $\dot{\gamma}_{\mathrm{hex}}\tau\approx257$
(below which the hexagonal state of the uncontrolled system is unstable) is given by 
\begin{equation}
\tau_{\mathrm{decay}}=\frac{\tau_c \eta_0}{m}\ln\left(\frac{m \dot{\gamma}_{\mathrm{init}} - \sigma_{0}}{m\dot{\gamma}_{\mathrm{hex}}-\sigma_{0}}\right).
\label{Control_an2}
\end{equation}

To estimate the reorganization time $\tau_{\mathrm{reorg}}$ (from the initial square into a hexagonal configuration), we assume that its dependence on 
$\dot\gamma_{\mathrm{init}}$ is analogous to that 
of the relaxation time $\tau_1$ introduced for the {\em uncontrolled} system [see eqn.~(\ref{intrinsic})]. 
Specifically, we make the ansatz
\begin{equation}
\tau_{\mathrm{reorg}}=\frac{a'}{|\dot{\gamma}_{\mathrm{init}}\tau-\dot{\gamma}_{\mathrm{hex}}\tau|^{b'}}.
\label{Control_an3}
\end{equation}
As stated above, a crucial assumption of our model is that the system can only reach 
the hexagonal state if $\tau_{\mathrm{reorg}}$ does not exceed $\tau_{\mathrm{decay}}$. Note that the latter involves (in fact, is proportional to) the time $\tau_c$. By
equating expressions (\ref{Control_an2}) and  (\ref{Control_an3}) for $\tau_{\mathrm{decay}}$ and $\tau_{\mathrm{reorg}}$, respectively, we can therefore find an expression
for the {\em minimal} control time, $\tau^{\mathrm{trans}}_c$, above which the system reaches the hexagonal state, that is
\begin{equation}
\tau_{c}^{\mathrm{trans}}=\frac{a' m}{|\dot{\gamma}_{\mathrm{init}}\tau-\dot{\gamma}_{\mathrm{hex}}\tau|^{b'}\eta_0\ln\left(\frac{m \dot{\gamma}_{\mathrm{init}} - \sigma_{0}}
{m\dot{\gamma}_{\mathrm{hex}}-\sigma_{0}}\right)}.
\label{Control_an4}
\end{equation}
Due to the square initial configuration, we set $m=\eta_0$ and $\sigma_{xz}(t)=m\dot\gamma(t)$ as defined in our {\em ansatz}. The remaining parameters  $a'$ and $b'$ are determined by fitting the numerical results for $\tau_c/\tau$ at the boundary [see Fig.~\ref{FIG:Heatmap}] to expression~(\ref{Control_an4}), yielding
$a'/\tau=54.127$ and $b'=1.503$. The resulting line $\tau_{c}^{\mathrm{trans}}(\dot\gamma_{\mathrm{init}})$ is included in Fig.~\ref{FIG:Heatmap}, showing that our estimate describes the transition between square and hexagonal states very well. 
\\
\begin{figure}
\includegraphics[width=0.9\linewidth]{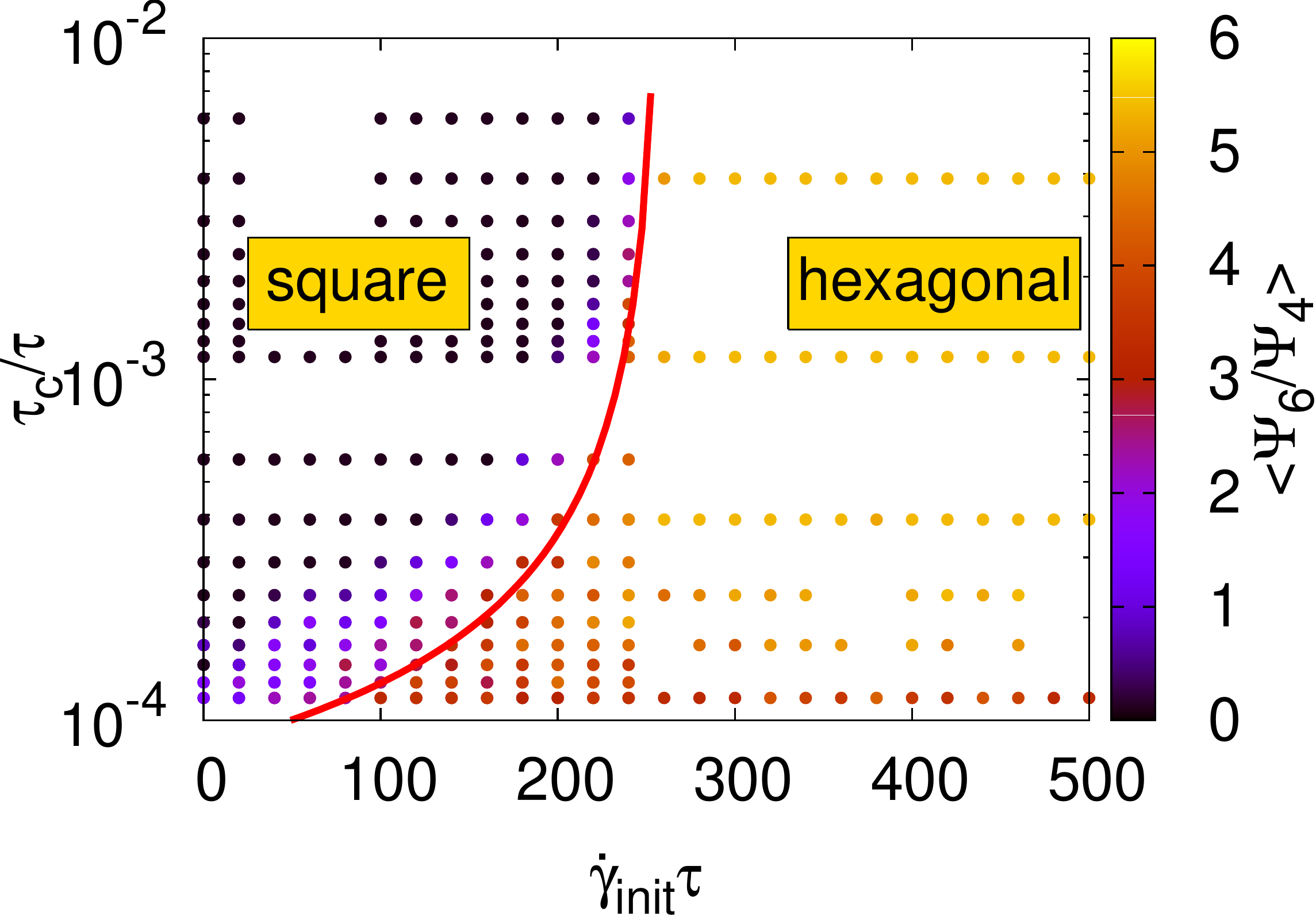}
\caption{(Color online) State diagram indicating long-time lattice structures. 
All simulations were started from the hexagonal initial structure and the imposed shear stress was set to $\sigma_0 d\tau^2=8$.} 
\label{FIG:Heatmap_hex}
\end{figure}

Similar considerations are possible, when we use a hexagonal initial lattice structure. Choosing then a {\em small} value of $\gamma_{\mathrm{init}}\tau$ we find that we can switch between hexagonal and square state. This is illustrated in Fig.~\ref{FIG:Heatmap_hex}. To obtain the corresponding transition values of $\tau_{c}$, we use the same strategy as before, but take a different {\em ansatz} for the stress. Specially, we set $\sigma_{\mathrm{xz}}(t)=n+m\dot{\gamma}(t)$ which approximately describe the flow curve in the hexagonal state of the uncontrolled system. From the results plotted in Fig.~\ref{FIG:Viscosity} we find $n=7.0477/d\tau^2$ and $m=0.0025$.  The analog of the  eqn.~(\ref{Control_an4}) then reads
\begin{equation}
\tau_{c}^{\mathrm{trans}}=\frac{a' m}{|\dot{\gamma}_{\mathrm{init}}\tau-\dot{\gamma}_{\mathrm{hex}}\tau|^{b'}\eta_0\ln\left(\frac{n+m \dot{\gamma}_{\mathrm{init}} - \sigma_{0}}
{n+m\dot{\gamma}_{\mathrm{hex}}-\sigma_{0}}\right)}
\label{Control_an5}
\end{equation}
with $a'/\tau=0.012$ and $b'=0.237$ [see Fig.~\ref{FIG:Heatmap_hex}].
The result is visualized in Fig.~\ref{FIG:Heatmap_hex}. Comparing the typical control timescales at the transition with those seen in Fig.~\ref{FIG:Heatmap} we find that  $\tau_{c}/\tau$ which is necessary to switch from square into the hexagonal state [see Fig.~\ref{FIG:Heatmap}] is about two orders of magnitude larger than switching from hexagonal into the square state. We suspect that this difference results from the differences of the slope of the shear stress in the square and hexagonal regimes. 

\section{Conclusions\label{SEC:CON}}
Using numerical simulation we have studied the complex dynamical behaviour of sheared colloidal films under a specific type of shear-stress control. Our approach involves relaxation
of the shear rate in a finite relaxation time $\tau_c$, until the instantaneous stress matches its desired value. 
This approach is inspired by rheological experiments \cite{Hu98} where the instantaneous shear rate as function of time can be measured.
Focusing on systems with multivalued flow curves (resulting from successive non-equilibrium transitions) we have found that, by tuning $\tau_c$ and the initial conditions, it is possible to {\em select} a specific dynamical state. In the present system these are either a state with square in-plane ordering and high viscosity, or a hexagonal state with low viscosity.
Therefore, our study suggests a way to stabilize states with desired rheological properties, particularly shear viscosities. Moreover, we have proposed a model which relates the transition values of $\tau_c$ to relevant {\em intrinsic} relaxation times under sudden change 
of $\dot\gamma$.

Although most of our results pertain to a colloidal bilayer, the fact that we found analogous results for trilayers suggests that the proposed technique can also be applied for systems with larger number of layers. In fact, we think that this method, after some minor adaptations such as consideration of the kinematic (and, possibly, also the non-local) contributions in eqn.~(\ref{stress}), should also be applicable and fruitful in bulk systems. Indeed, we expect 
the method to allow for state selection in {\em any} shear-driven system with multivalued flow curve. For example, in an earlier study we have used an analogous approach (based, however, on continuum equations) to select states and even suppress chaos in shear-driven nematic liquid crystals \cite{Hess10}. It therefore seems safe to assume
that the capabilities of the present scheme are quite wide. For colloidal layers one may envision, e.g., stabilization of {\em time-dependent structures} such as oscillatory density excitation, 
which may have profound implications for lubrication properties \cite{Vanossi12}.

Finally, our findings are quite intriguing in the broader context of manipulating nonlinear systems by feedback control.  In our case, the feedback character stems from the fact that the stress control involves the configuration-dependent instantaneous stress. Mathematically, this scheme can be viewed as feedback control with exponentially distributed time-delay \cite{Hoevel06} (as can be seen by formally integrating eqn.~(\ref{Control}) and inserting it into eqn.~(\ref{EQ:Eqmot})). Similar schemes are used to stabilize dynamical patterns in laser networks \cite{Juengling12}, neural systems \cite{Song13}, and more generally, coupled
oscillator systems \cite{Kyrychko11}. The implications of these connections
are yet to be explored.

\section*{Appendix A: Stability of the feedback controlled system \label{SEC:APPB}}
In this appendix we investigate the stability of the solutions of eqn.~(\ref{Control}). Specifically, we consider the impact of small variations
of the shear rate from its steady state value $\dot\gamma_0$ related to the imposed stress $\sigma_0$.
Expanding the right side of eqn.~(\ref{Control}) with respect to the difference $\dot\gamma-\dot\gamma_0$ yields
\begin{equation}
\frac{d}{dt}\dot\gamma \approx \frac{\sigma_0-\sigma_{\mathrm{xz}}(\dot \gamma_0, t)}{\tau_c \eta_0}- \frac{\partial \sigma_{\mathrm{xz}}(\dot \gamma, t)}{\tau_c \eta_0 \partial \dot \gamma}  \bigg \vert _{\dot \gamma_0} (\dot \gamma-\dot \gamma_0) +O(\dot \gamma^2).
\label{eqn:appa1}
\end{equation}
For long times we expect the first term on the right side of eqn.~(\ref{eqn:appa1}) to vanish, since
$\sigma_{\mathrm{xz}}(\dot \gamma_0, t)\rightarrow\sigma_0$. To linear order, eqn.~(\ref{eqn:appa1}) then reduces to
\begin{equation}
\frac{d}{dt}\dot\gamma \approx - \frac{1}{\tau_c \eta_0}  \frac{\partial \sigma_{\mathrm{xz}}(\dot \gamma, t)}{\partial \dot \gamma}  \bigg \vert _{\dot \gamma_0} (\dot \gamma-\dot \gamma_0) +O(\dot \gamma^2).
\label{eqn:appa2}
\end{equation}
Noting that the values of $\tau_c$ and $\eta_0$ are both positive, we can follow that 
the feedback controlled shear rate approaches a steady-state value only if
\begin{equation}
\frac{\partial \sigma_{xz}}{\partial \dot \gamma}>0.
\end{equation}
This corresponds to the usual criterion of mechanical stability \cite{Tanner}. 
\section*{Appendix B: Strain-stress relation under constant shear rate \label{SEC:APPA}}
\begin{figure}[h]
\includegraphics[width=0.8\linewidth]{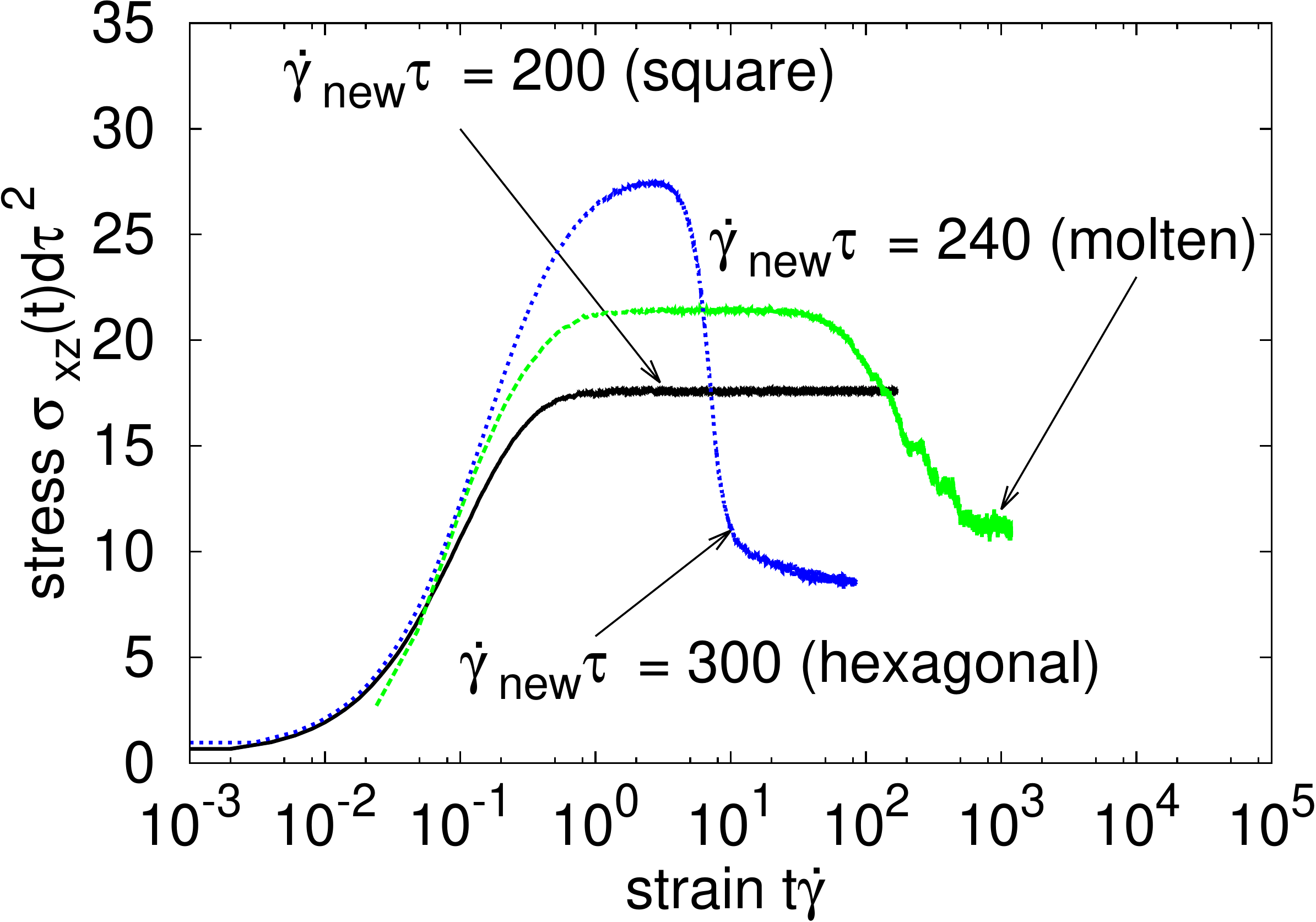}
\caption{(Color online) Stress-strain relations in the colloidal bilayer for different shear rates, starting from the equilibrium (square) configuration. }
\label{FIG:Strain}
\end{figure}

In this Appendix we present results for stress-strain relations at different fixed values of $\dot\gamma$. These can be obtained from the data shown in Fig.~\ref{FIG:Relaxation1} by rescaling the time axis with the applied shear rate. Numerical results are shown
in Fig. \ref{FIG:Strain}.  
Similar to the stress-time relations shown in Fig.~\ref{FIG:Relaxation1}, one observes simple, monotonic behavior for the case 
$\dot\gamma_{\mathrm{new}}=200$ (quadratic regime), while the curves for larger shear rates display pronounced 
stress overshoots).
The width of these overshoots is largest at $\dot\gamma_{\mathrm{new}}=240$, where the system is in the molten state. This is consistent with the appearance of a particularly large intrinsic relaxtion time as discussed in sec. \ref{SEC:TS}. 
\\
\\
{\large{\textbf{Acknowledgments}}}\\
This work was supported by the Deutsche Forschungsgemeinschaft through SFB 910 (project B2).





\footnotesize{
\bibliography{rsc} 
\bibliographystyle{rsc} 
}

\end{document}